\documentclass{aa}
\usepackage{graphicx}
\usepackage{txfonts}
\usepackage{psfig}
\begin{document}

%\thesaurus{03       % Galaxies
%          (03.04.1; % Clusters of galaxies,
%           07.17.1) % radio galaxies  }

\title{The spiral galaxy M\,33 mapped in the FIR by ISOPHOT:} 
\subtitle{A spatially resolved study of the warm and cold dust. 
\thanks{Based on observations with the Infrared Space Observatory ISO. 
ISO is an ESA project with instruments funded by ESA Member States 
(especially the PI countries France, Germany, The Netherlands and the UK) 
and with the participation of ISAS and NASA.}} 

\author{H. Hippelein\inst{1} \and M. Haas\inst{1} \and 
R.J. Tuffs\inst{2} \and D. Lemke\inst{1} \and 
M. Stickel\inst{1} \and U. Klaas\inst{1} \and H.J. V\"olk\inst{2}} 

\institute{ Max--Planck--Institut f\"ur Astronomie, K\"onigstuhl 17, 
            D-69117 Heidelberg, Germany 
       \and Max--Planck--Institut f\"ur Kernphysik, Saupfercheckweg, 
            D-69117 Heidelberg, Germany} 

\date{Received  2003; Accepted } 

%\offprints{H. Hippelein} 

\abstract{ 
The Sc galaxy M\,33 has been mapped with ISOPHOT in the far-infrared, at 
60, 100, and 170\,$\mu$m. The spatial resolution of these FIR maps 
allows the separation of spiral arms and interarm regions and the 
isolation of a large number of star-forming regions. 
The spectral energy distribution in the FIR indicates a superposition of 
two components, a warm one originating from dust at $\sim45$K, and a cold 
one, at $\sim16$K. The warm 
component is concentrated towards the spiral arms and the star-forming 
regions, and is likely heated by the UV radiation from OB stars. The 
cold component is more smoothly distributed over the disk, and heated by 
the diffuse interstellar radiation. 
For the about 60 star-forming regions detected the H$\alpha$/FIR flux ratio 
increases significantly with the distance from the galaxy center, probably 
due to decreasing extinction. 
An anti-correlation of $F_{Ha} / F_{60}$ with $F_{170}$ suggests 
the intrinsic extinction to be related to the cold dust surface brightness 
according to $A_V / S_{170} \sim 0.03$\,mag\,MJy$^{-1}$\,sr. 
For the total galaxy the star formation rate (SFR) derived from the FIR is in 
agreement with that derived from the de-extincted H$\alpha$ emission. 
For individual star-forming regions, a consistency between SFRs derived 
from the optical and from the FIR requires only a fraction of the 
UV radiation to be absorbed locally. 
The individual star-forming regions also show a local radio-FIR correlation. 
This local correlation is, however, due to quite different 
components than to those that lead to 
the well-known global radio-FIR correlation for entire galaxies. 
} 
\titlerunning{Dust in M\,33} 
\authorrunning{Hippelein et al.} 
\maketitle

\section{Introduction}    %1 

The sensitivity of ISO (Kessler et al. 1996) and its spectral coverage 
extending to 200\,${\mu}$m made it the first observatory capable of 
measuring the bulk of starlight absorbed by dust in galaxies. 
Recent FIR studies of normal galaxies (e.g. Haas et al. 1998 for M~31; see 
also Tuffs \& Popescu 2002 for a review), based on data taken 
with the ISOPHOT instrument (Lemke et al. 1996) onboard ISO, have proven 
the existence of a cold dust component with a temperature below 20\,K, in 
addition to the warm component near $T\sim50$K. These  components were 
originally proposed by Chini et al. (1986), based on sub-mm observations. 
Statistically significant evidence for the existence of a cold dust component 
was also inferred from the ISOPHOT Serendipity Survey (Stickel et al. 2000), 
and from the analysis of a complete volume- and luminosity limited sample of 
late-type Virgo Cluster galaxies - the ISOPHOT Virgo Cluster Deep Sample 
(Tuffs et al. 2002, Popescu et al. 2002). Even 
ultra-luminous IR galaxies show evidence for a cold dust component (Klaas et 
al. 2001), though this does not constitute the bulk of the FIR emission, as in 
the case of normal galaxies. 
This cold component arises from grains which are in weaker radiation fields, 
either in the outer optically thin regions of the galaxy disks or because they 
are shielded from radiation by optical depth effects (Tuffs \& Popescu 2002). 

Though the existence of the warm and cold components were unambiguously 
inferred from the statistical studies mentioned above, the spatial 
distributions of these components can only be studied by the analysis of 
resolved nearby galaxies. Such an analysis was done for the ISOPHOT map of M~31 
by Haas et al. (1998) and Schmidtobreick et al. (2000), who resolved several dozens 
of bright knots and separated their SEDs in cold and warm dust components. 
While the FIR emission in M\,31, a Sab galaxy seen at high inclination, originates 
mainly from a ring-like structure rather than from spiral arms, M\,33, 
a Sc galaxy with well distinct spiral arms, and seen almost face on, seems 
to be an ideal target to study the spatial distribution of dust and its heating 
sources in spiral galaxies.  Due to its distance of only 
830\,kpc, its scale of 0.24\,kpc\,arcmin$^{-1}$ is sufficient to allow a 
separation of its constituents (nucleus, H\,{\sc ii} regions, spiral 
arms, and interarm regions) with ISOPHOT even at 170\,$\mu$m. 

We will make use of the higher spatial resolution and the extended spectral 
coverage (to longer FIR wavelengths) of ISOPHOT as compared with IRAS to 
disentangle the warm and cold dust 
components in M\,33. The spatial distributions of these dust components will 
be compared with the distribution of the optical emission. The FIR emission 
from individual HII regions will be correlated with the corresponding radio 
emission. In this paper we will also study the relation between FIR emission 
and star formation rate, as well as the heating of the two dust emission 
components and their effects on extinction.

\section{Observations and data reduction} 

\subsection{FIR data} 

M\,33 was mapped with ISOPHOT (Lemke et al. 1996) in the raster scan mode, AOT-P32, 
at three wavelengths, 60, 100, and 170\,$\mu$m, using the C100 and C200 detector arrays. 
The raster step sizes in spacecraft coordinates were 69\arcsec$\,\times\,$92\arcsec\ 
for 60 and 100\,$\mu$m and 92\arcsec$\,\times\,$92\arcsec\ at 170\,$\mu$m. 
Com\-bined with the detector geometry and the motion of the focal plane chopper, 
this yiel\-ded a sky sampling of 15\arcsec$\,\times\,$23\arcsec\ 
and 31\arcsec$\,\times\,$92\arcsec\, respectively. 
Since the linear extension of M33 is too large to be coverable by one 
raster map with the desired resolution, two partial maps had to be 
done in each filter (North and South) and concatenated. 
With the partial Northern and Southern P32 maps mosaiced together 
according to their center positions and satellite roll angles, a total area of 
48\arcmin\,$\times$\,32\arcmin\ was covered, including the entire disk of M33, and 
a surrounding background area to allow for proper sky subtraction. 
The orientation of this map is roughly along the major axis of the galaxy. 

At 170$\mu$m the central area of 29\arcmin $\times$\,26\arcmin\ size was also mapped in 
the AOT-P22 raster staring mode, which almost is free from 
detector transient effects due to the relatively long integrating time of 10 seconds 
per raster position. The P32 and P22 maps were merged to a common 
one by using equal weights for both maps in the bright and medium bright areas and 
zero-weight for the P22 map for areas where the surface brightness 
falls below 13\,MJy/sr. 

The data reduction was done using the ISOPHOT Interactive Analysis software package 
PIA V7 (Gabriel et al. 1998), together with the calibration data set V4.0 (Laureijs 
et al. 2000). A special data handling was required due to signal transient effects 
of the infrared detectors, which is described in sect. 2.2. 
The absolute flux calibration was performed with measurements of the internal 
thermal fine calibration source on board, before and after each map. 
We estimate the error of this flux calibration to be about $\pm20\%$. An additional 
uncertainty is introduced by the detector transient effects in the P32 observing 
mode, which generally leads to too low signals for the FIR knots, and which we have 
only corrected to a certain degree (see below). At 60 and 100\,$\mu$m, where 
this effect is most critical, we estimate the total error to be $\pm30\%$. 
At 170\,$\mu$m the transient effect is less severe, and the total 
error is of the order of $\pm20\%$.

\subsection{Handling of transient detector effects} 

As mentioned above, a difficulty in the reduction of P32 data are signal transient 
effects of the infrared detectors, leading to small lags of the maxima positions 
for bright point-like sources along the scan direction, and to a 
lowering of the maxima and filling up of the minima. It also led  to 
ghost images of bright sources at distances of the chopper throw in the 
scan direction. 
Based on the principles described by Acosta-Pulido et al. (2000) we adopted 
a correction routine to reduce the transient effects:

\begin{figure*} 
\centerline{ 
\vbox{ 
\hbox{ 
\psfig{figure=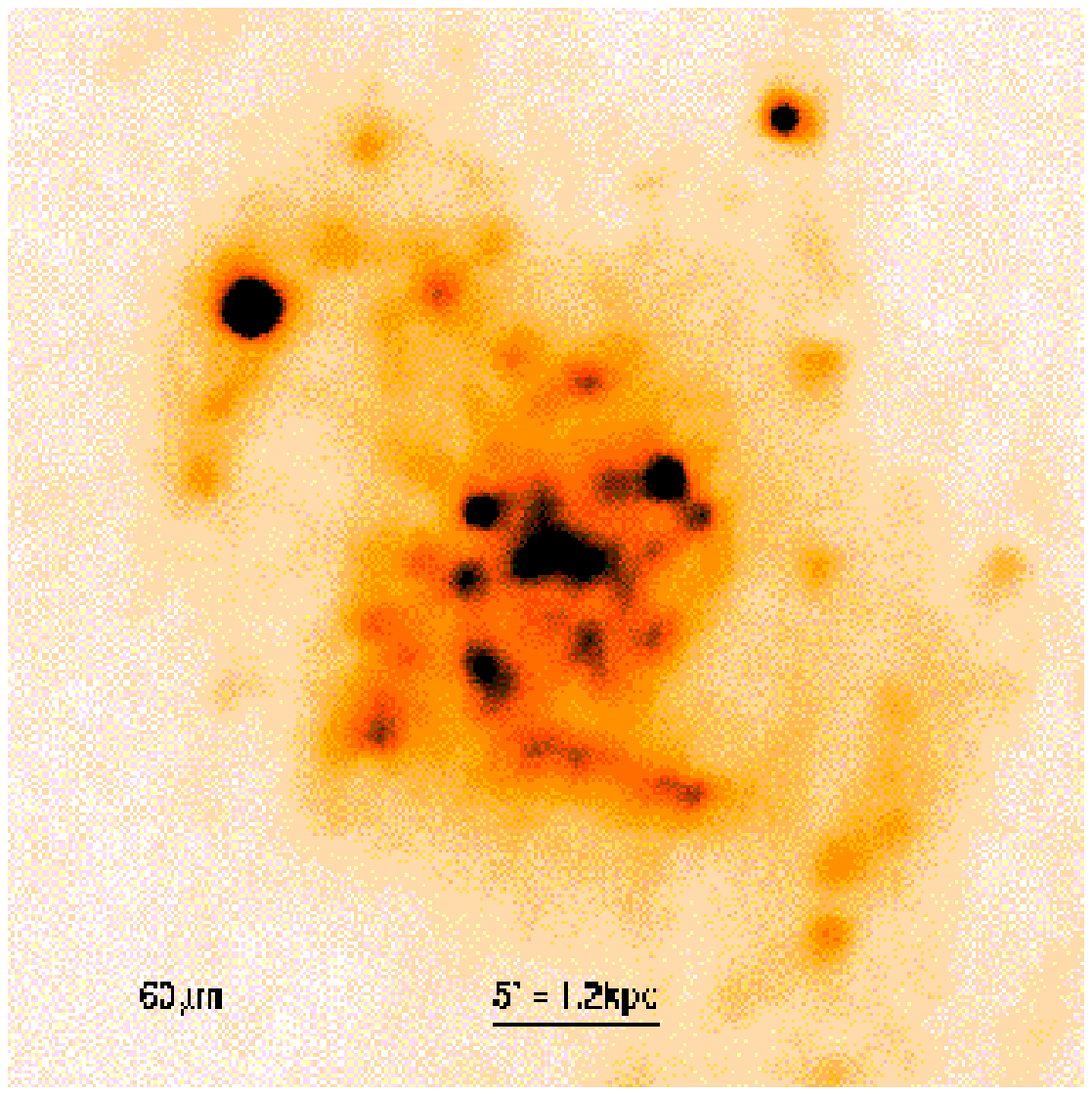,clip=t,width=8.8cm}\hspace{3.mm} 
\psfig{figure=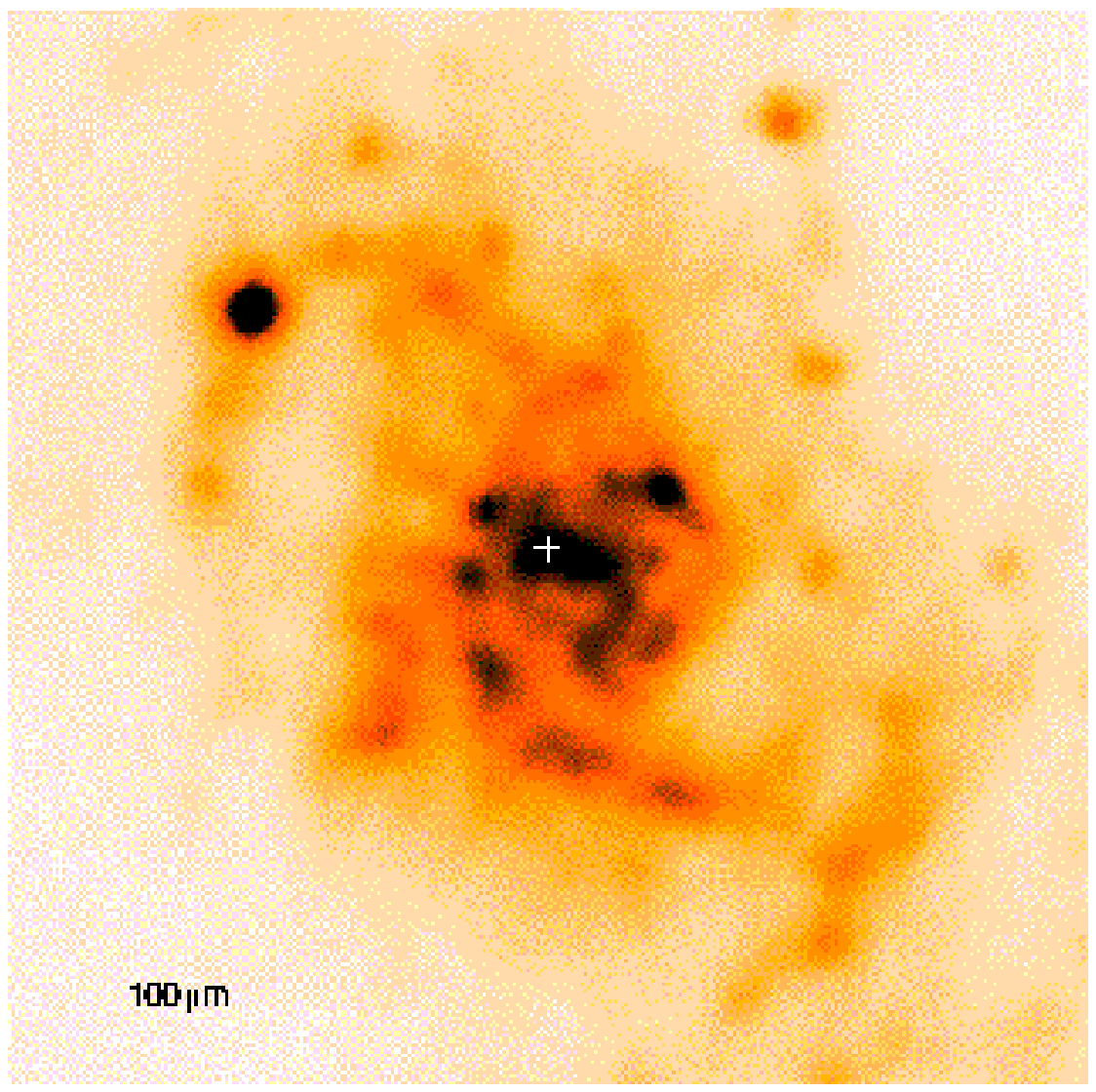,clip=t,width=8.8cm}}\vspace{3.mm} 
\hbox{ 
\psfig{figure=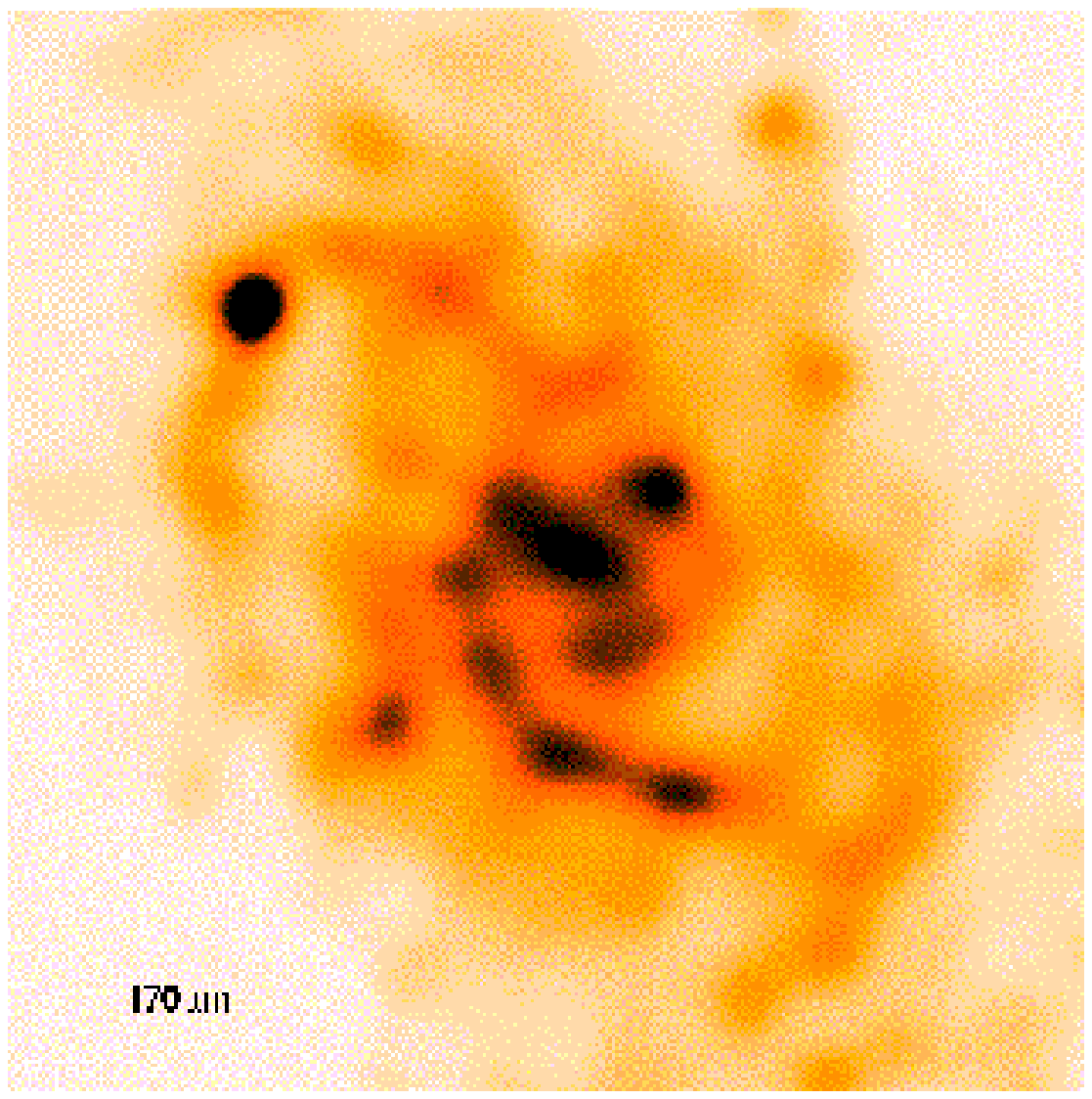,clip=t,width=8.8cm}\hspace{3.mm}  
\psfig{figure=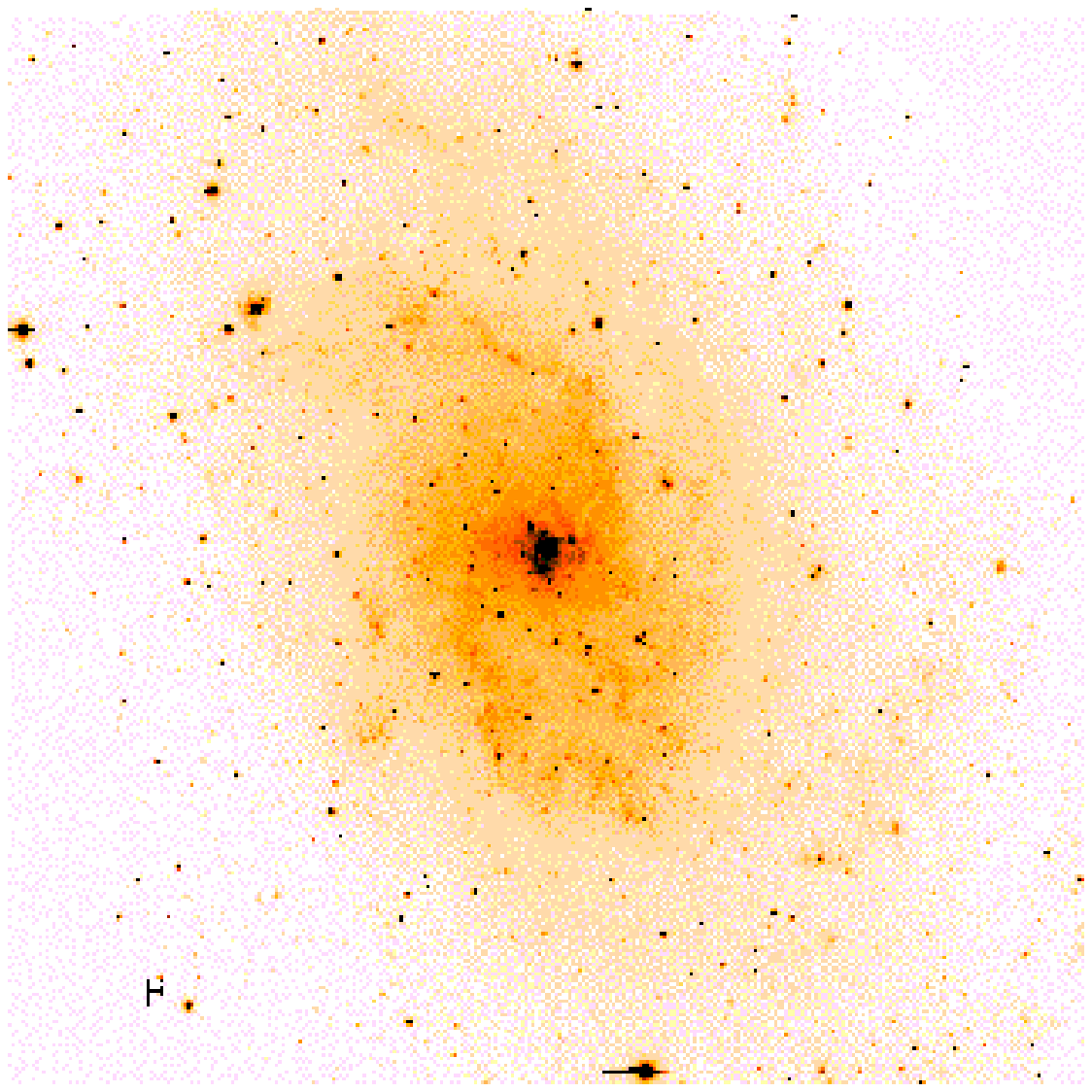,clip=t,width=8.8cm}}}} 
\caption[]{FIR maps of M33 (area 33\arcmin$\times$33\arcmin, North is up, 
East is left) at 60 (left top), 
100 (right top), and 170\,$\mu$m (left bottom). The 170\,$\mu$m map is restored to a 
spatial resolution of 100\arcsec, while the resolutions at 60 and 100\,$\mu$m are 
60\arcsec\ and 80\arcsec\, respectively. For comparison, an R image of the galaxy 
is also shown. (The full resolution images of Fig.1 can be obtained by anonymous ftp 
from ftp.mpia.de /pub/iso-arch).} 
\end{figure*}

First, separate maps were constructed for each detector pixel, nine for the 60 and 
100$\mu$m maps, which were observed with the C100 detector array. and four in the 
case of the 170$\mu$m map, which was observed with the C200 array. 
Transient effect were reduced by adding to the signal in each pixel 30\% of the 
difference to the signal of the preceeding pixel during the scan. This procedure led 
to a steepening of the edges of bright sources and improved the coincidence in the 
forth and back scans. 
It also led to a significant reduction of the ghosts from bright point like sources 
at the map pixels located in chopper throw separation (here, however, a factor of 
only 10\% was applied to the difference signal). 

Second, the separate pixel maps were scaled to a common level and superimposed. 
Due to a displacement of the raster legs by half of the array size and a chopper 
step of 15\arcsec\ the pixel size of the resulting image was 15\arcsec$\times$23\arcsec\ 
(in the cases of 60 and 100$\mu$m). 
Each of these pixels was seen by 3 detector pixels in the one raster leg in forth 
direction and in the adjacent raster leg in opposite direction by 3 other detector pixels. 
From the 6 detector signals in each pixel the median signal was derived, and an 
iterative $\kappa$-$\sigma$ clipping was applied toin the separate pixel maps and to correct them. This also leads to a spatial resolution 
of 60\arcsec\ at 60$\mu$m, of 80\arcsec\ at 100$\mu$m, and of 120\arcsec\ at 
170$\mu$m. 
It was found that each scan had a slight positive slope, recognizable by small 
offsets at the edges of the forth and back scans. This was corrected by 
a fine adjustment of the scan slopes until a smooth background was achieved. 
The background level (from foreground cirrus and zodiacal light) was then determined 
from the outermost map areas and subtracted.

\subsection{Optical observations} 

For studying the correlation of optical emission with the detailed 
ISOPHOT maps, M\,33 was observed in H$\alpha$ 
with the focal reducer at the 2.2\,m telescope 
on Calar Alto, Spain. 
The detector was a 2k$\times$2k SITe chip used in a rebinned mode, to provide 
a scale of 1\farcs06~pixel$^{-1}$, thus covering a field of 15\arcmin diameter. 
The photometric conditions were good. The seeing of 2\farcs5 was 
adequate for this study. 
For isolating the H$\alpha$ line emission a 5.2nm wide filter, centered 
at 657.0nm was used. For continuum subtraction we used a broad band red 
filter, centered at 650nm. To cover the whole galaxy, a mosaic of 13 positions 
was obtained. The single maps were corrected for the image distortion of 
the focal reducer (with flux conservation) and then aligned using 
stellar DSS positions and merged to an H$\alpha$ map of 
55\arcmin$\times$35\arcmin\ size. 
The flux calibration for the H$\alpha$ map was done using the total line 
flux value given by Devereux et al. (1997) and using the H$\alpha$ line 
fluxes of the prominent H\,{\sc ii} regions tabulated in Hippelein (1986), 
assuming a mean relative intensity of the [NII] lines of 20$\%$ (McCall et 
al. 1985).

\section{Results}                    % 3

In this section we will first present the general morphology of M~33, then we 
present the FIR photometry of discrete sources and their FIR spectral 
properties. 

\subsection{Morphology}              % 3.1

\begin{figure}                       % Fig 2 
\centerline{\psfig{figure=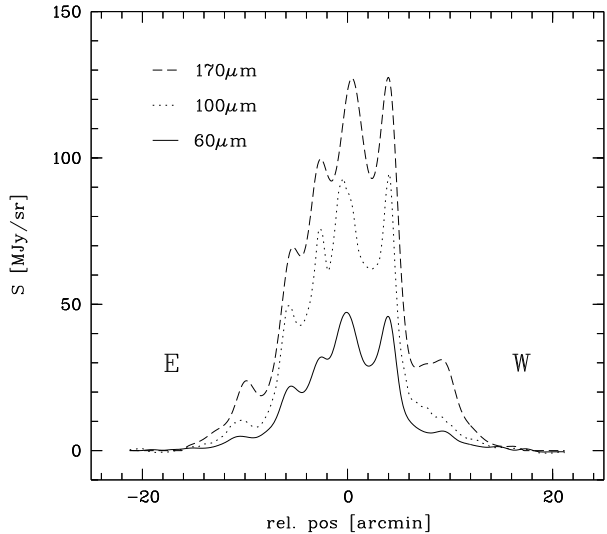,clip=t,width=8.2cm}} 
\caption[]{Profiles through the center of M33 along position angle 115\degr\ 
for the three FIR wavelengths. Here, the 60$\mu$m map is convolved with a Gaussian 
of 60\arcsec\ FWHM to achieve a similar spatial resolution as the 100$\mu$m 
(80\arcsec\ FWHM), and the 170$\mu$m profile is taken from the Lucy-Richardson 
restored map. } 
\end{figure}

Fig. 1 presents the maps obtained at the three FIR wavelengths, rebinned to a 
pixel size of 5\arcsec. 
The spatial resolutions are 60\arcsec\ at 60$\mu$m and 80\arcsec\ at 
100$\mu$m, as derived from the profile of IC133, the bright region in the NW 
edge. 
In the case of 170$\mu$m the spatial resolution of $\approx$120\arcsec, 
as measured from the profile of NGC604 (IC133 is too faint at 170$\mu$m), 
was improved to 100\arcsec\ by deconvolving the original map with a PSF of 
120\arcsec. 

All three FIR maps clearly show the spiral arm structure of the galaxy 
and a large number of distinct sources. In addition to the spiral 
structure, an extended underlying component can be recognized, which 
becomes more significant with increasing wavelength. This is illustrated by 
the intensity profiles in Fig. 2,  showing cuts through the three 
FIR maps at a position angle of 115\degr, 
along the minor axis. The structure of the profiles is identical for the 
three filters, except that at the edge of the galaxy the 170$\mu$m profile 
seems to decay slower than the others. 

The total flux density of M33 is 560Jy, 1250Jy, and 2200Jy at 60, 100, and 
170$\mu$m, respectively. 

\subsection{FIR photometry of discrete sources}   % 3.2

About sixty sources were isolated from the FIR maps shown in Fig. 1. 
This number is large enough to allow statistical studies for the FIR properties 
of extragalactic star-forming regions. 
Aperture photometry was done for these sources, with aperture 
diameters between 100\arcsec\ for small regions, up to 260\arcsec\ for 
the giant star-forming region NGC\,604. 
Since in the crowded areas the diffuse emission from M\,33 cannot be 
unambigously determined, and since the introduction of a model for this 
component would be rather subjective and lead to arbitrary results, the 
photometry is performed without a subtraction of the diffuse emission.

\begin{figure*}[ht]              % Fig 3 
\centerline{\psfig{figure=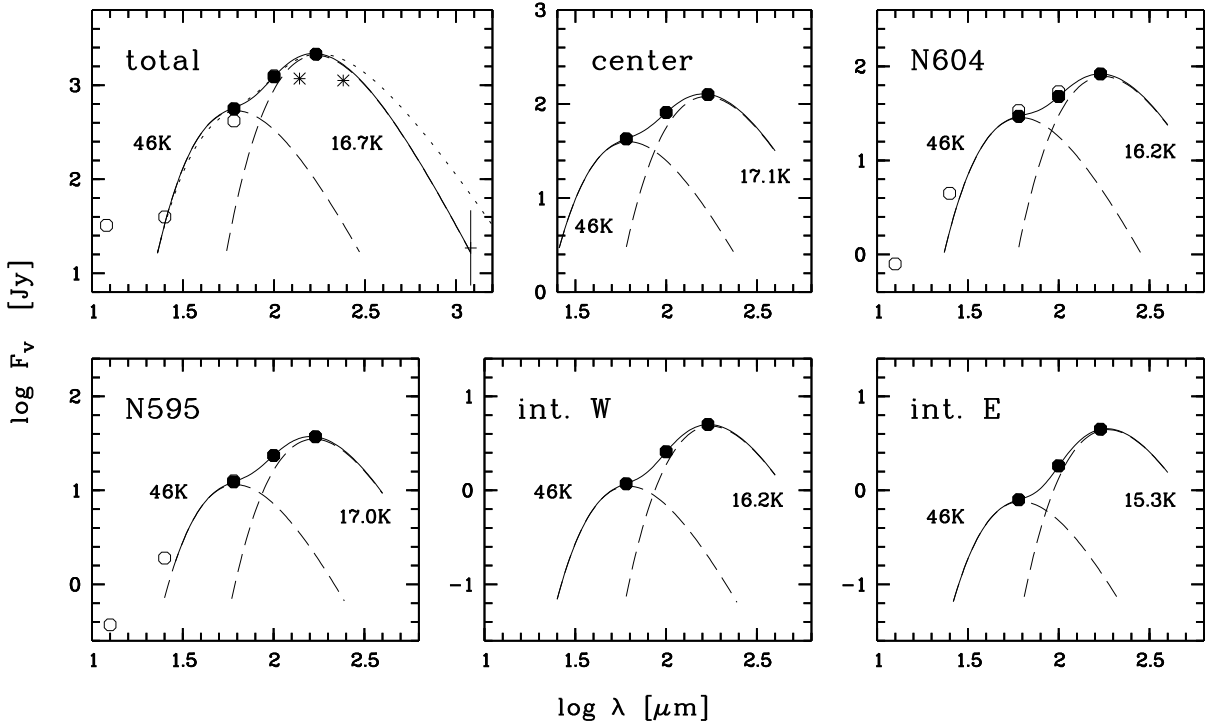,clip=t,width=17cm}} 
\caption[]{Spectral energy distribution for total M33 (top left), for the central area 
(top center), for two prominent H\,{\sc ii} regions (NGC\,604, NGC\,595), and 
for two areas in the interarm regime, W int. (offset 6\arcmin\ W, 5\arcmin\ S 
from the nucleus) and int. E (8\arcmin\ E from the nucleus). 
Dots are ISOPHOT data, circles stand for IRAS, asterisks for COBE/DIRBE data, 
and the plus with the large error bar at 1200$\mu$m is derived from flux density 
ratios as estimated by Chini et al., 1986). Two-component blackbody fits 
for temperatures of T$_{warm}=46.0$\,K, T$_{cold}\sim16.5$\,K, respectively, 
and with emissivity $\beta \propto \lambda^{-2}$ are overlaid as dashed and solid lines. 
For comparison, the dotted line shows a fit for $\beta \propto \lambda^{-1}$ with 
T$_{warm}=52.0$\,K, T$_{cold}=19$\,K.} 
\end{figure*} 

From IRAS maps of M\,33, Rice et al. (1990) localized a number of H\,{\sc ii} complexes 
and derived flux densities at 60 and 100$\mu$m. Their 60$\mu$m flux densities (Table 1) 
are in reasonable agreement with ours, while their 100$\mu$m data are 
generally higher, probably due to the inferior spatial resolution of IRAS.

\begin{table}[ht!] 
\caption{Comparison with IRAS results of Rice et al. (1990). $\Phi$ are the aperture 
diameters used 
for the ISO photometry, $F_{60}$ stands for flux density $F_{\nu}$ at 60\,$\mu$m in Jy, etc. } 
\begin{center} 
\renewcommand{\arraystretch}{0.97} 
\footnotesize 
\begin{tabular} 
{   l        c       r          r         r        r     } 
\hline 
\noalign{\smallskip} 
 Name   & $\Phi$ &$F_{60}$&$F_{100}$& $F_{60}$&$F_{100}$\\ 
        & \multicolumn{3}{c}{ISOPHOT} & 
                       \multicolumn{2}{c}{IRAS} \\[1mm] 
                               \noalign{\hrule} \\[-2mm] 
 total  &        & 560   &  1250  &  420   & 1260   \\ 
 B651   &  100\arcsec  &  1.0  &   1.9  &   1.7  &   4.5  \\ 
 IC133  &  140\arcsec  &  6.3  &   7.3  &  11.8  &  10.7  \\ 
 IK60   &  120\arcsec  &  2.4  &   4.6  &   2.4  &   ..   \\ 
 NGC604 &  260\arcsec  & 29.5  &  46.4  &  33.6  &  54.5  \\ 
 IC131  &  140\arcsec  &  3.2  &   6.4  &   2.7  &   3.2: \\ 
 NGC595 &  140\arcsec  & 13.0  &  22.9  &  12.8  &   ..   \\ 
 B220/1 &  100\arcsec  &  2.4  &   5.4  &   1.9  &   ..   \\ 
 B255/7 &  100\arcsec  &  1.7  &   4.1  &   1.6  &   ..   \\ 
 B248-51&  120\arcsec  &  2.6  &   5.5  &   2.7  &   8.3  \\ 
\end{tabular} 
\end{center} 
\end{table}

\subsection{Spectral energy distribution}    % 3.3

Fig. 3 depicts the SEDs as derived from the photometry for a number of selected 
areas. In the cases of the total galaxy and of NGC\,604 and NGC\,595 the IRAS 
(Rice et al. 1990) points are overplotted. The COBE/DIRBE (Odenwald et al. 1998) data 
overplotted in the SED for the total galaxy are somewhat lower. The explanation for 
this difference is that M33 is situated in a small hole in the cirrus foreground, 
and the off-galaxy background measured by COBE/DIRBE with its large beam in the 
neighbourhood is probably too high, leading to lower flux densities for the galaxy.

In all cases (Fig.\,3) it is obvious from the distribution of the three FIR 
data points that the SEDs cannot be represented by a single blackbody, but need 
a combination of two blackbody emission components. 
We associate these two emissions with the two main morphological components 
measured directly in this paper, namely the localised (mainly warm) and the diffuse 
(mainly cold) component. 
Two blackbodies have four degrees 
of freedom and as there are only three data points, no unique combination can be 
derived. In the cases of M\,33(total) and NGC\,604 however, the IRAS data 
point at 25$\mu$m puts an upper limit to the temperature of the warm blackbody 
component. 

At the long wavelength end Chini et al. (1986) found for Sb and Sc type galaxies 
observed by IRAS an average ratio S$_{1200{\mu}m}/$S$_{60{\mu}m}\sim 0.034$, with 
a scatter of a factor 2.5. Adopting the same ratio for M33 we insert a 1200\,$\mu$m 
data point into the M33 SED. 

Attributing 30$\%$ of the signal at 25\,$\mu$m to hot dust, the fit with two 
modified blackbody functions (assuming an emissivity of $\beta \propto \lambda^{-2}$) to the 
four data points between 25 and 170\,$\mu$m yields T$_{warm}=46.0\pm0.8$\,K, and 
T$_{cold}=16.7\pm0.5$\,K. As seen in the plot the cold blackbody curve comes very 
close to the 1200\,$\mu$m point estimated above. 
Adopting the temperature of the warm component to be rather constant over the 
galaxy, we keep T$_{warm}=46$\,K fixed for fitting the cold component in the 
other areas. The resulting T$_{cold}$ varies only slightly, taking values between 
15.3 and 17.1\,K, 
as demonstrated in the figure, with error bars of typically $\pm0.5$\,K. 
This indicates that (coincidentially) the temperature of the diffuse cold dust 
component is rather close to the temperature of the localised cold dust component. 
If one does the same for $\beta \propto \lambda^{-1}$, one would get 
T$_{warm}=52.5\pm1.0$\,K, and T$_{cold}=19.4\pm0.7$\,K for M33$_{total}$, with a 
submm value a factor of 3 above the adopted one (dotted curve), while the variation 
of T$_{cold}$ was again rather small over the field.  

Therefore, we interpret the spatial variation of the flux density ratio 
$F_{warm}/F_{cold}$ as being due to a variation of their amplitudes, 
rather than being due to a change of their temperatures, which seem to remain 
almost constant over the galaxy. 
Since the warm and the cold component peak very close to 
60 and 170\,$\mu$m, respectively, the ISOPHOT 60 and 170\,$\mu$m flux 
densities can be used as tracers of the warm and the cold dust.

\section{Discussion}                            % 4

\subsection{Spatial distribution of the warm dust}   % 4.1

In Fig. 4, the ratios $F_{60}/F_{170}$ for all FIR knots are plotted versus 
the deprojected radial distance from the center of the galaxy. For the 
deprojection an inclination of 56\degr\ and a position angle of PA$= 23$\degr\ 
was assumed for the galaxy (Tully 1988). 
In addition a number of positions in the interarm regions (open circles) were 
selected. 
Firstly, it is obvious that the flux density ratio for the interarm areas 
is lower than that of the bright regions at the same radius. 
Secondly, both the ratios in the interarm and bright regions decrease with 
increasing radial distance, except for the bright H\,{\sc ii} regions NGC604 
and IC133. 
In the case of the interarm regions, this decrease can be explained by a 
decrease in the 60/170 micron colour temperature of the diffuse emission 
of M\,33. In the case of the bright star-forming regions, the decrease 
in the $F_{60}/F_{170}$ ratio is attributed to an increasing relative 
contribution of the diffuse 170$\mu$m emission from the disk in the 
fainter, outer parts. Very bright regions, such as NGC604, but also those in 
the central area are almost unaffected by this diffuse disk component. 
A special case is IC133 due to its extraordinarily high F$_{60}$ value. 
Schulman \& Bregman (1995) found it associated with a bright X-ray source 
located in a hole in the H{\sc i} layer of the galaxy, indicating energetic 
stellar winds and supernovae from massive stars. The average $F_{60}/F_{170}$ 
ratio for the 20 bright star-forming regions is 0.32, with a dispersion 
of 0.05, and 0.165 for the interarm regions with a scatter of 0.03. 

The observed FIR brightness distribution is a superposition of localized 
emission from star-forming regions and emission from a diffuse component. The 
localized FIR emission is mainly powered by UV 
photons, whereas the diffuse FIR emission is powered by both UV and optical 
photons. 
Models of the UV-optical/FIR-submm SEDs which self consistently calculate a 
continuous distribution of dust temperatures based on radiative-transfer 
calculations (see Popescu \& Tuffs 2002a) predict that, for $\lambda<100$\,$\mu$m, 
the FIR emission is dominated by the localized component. These 
models also predict that at longer wavelengths most emission originates from 
the diffuse disk component, and that both the localized and the diffuse dust 
emissions are predominantly powered by the UV light. 

We will try to separate the localized FIR emission (from star-forming regions) 
from the diffuse FIR emission component. 
While for central distances larger than $r=8$\arcmin\ the interarm areas are 
wide enough to allow an estimate of the diffuse radiation from the disk, 
this is not possible in the inner area of the galaxy. 

Two simple approaches are performed. 
Firstly,  assuming an exponential disk with a central flux density of 25\,MJy/sr 
at 60\,$\mu$m and a scale length of 6\arcmin\ (1.44\,kpc) for the diffuse 
emission component would yield a residual map for the localized emission with 
a level in the inter-arm regions close to zero. The contribution of the diffuse 
disk to the total 60$\mu$m radiation would be 42\%. 
This number agrees with the model for the Sb galaxy NGC\,891 by Popescu et al. 
(2000) for which the contribution of the diffuse emission to the total is 
about 40$\%$ at 60\,$\mu$m. 

A second approach to obtain the contribution of the star-forming regions at 60 
microns is to make a linear combination of the 60 and 170\,$\mu$m maps. 
If we denote the diffuse emission component at 60 and 170 micron by 
$F_{60}^{d}$ and $F_{170}^{d}$, respectively, and correspondingly the localized 
emission components by $F_{60}^{l}$ and $F_{170}^{l}$, then the arm and 
interarm colour ratios obtained previously can be written as 
$f^{l} =F_{60}^{l}/F_{170}^{l}$ and $f^{d} =F_{60}^{d}/F_{170}^{d}$, 
respectively. In this case the localized emission component at 60 micron is 
given by the following equation, with $F_{60}$ and $F_{170}$ in Jy: 

\begin{eqnarray} 
F_{60}^{l} = \frac{F_{60} - f^{d} \times F_{170}}{1 -f^{d}/f^{l}} 
\end{eqnarray} 
For $f^{d}=0.165$ and $f^{l}=0.32$, we find that $80\%$ of the emission at 60 
micron comes from the localized component. So only $20\%$ of the emission is 
diffuse at this wavelength. The $20\%$ value is lower than that obtained with 
the first method. However this is to be expected, since here we neglected the 
radial dependence of the $f^{d}$ and $f^{l}$ factors. 
The result in Fig. 5 shows the star-forming regions well isolated against a 
zero-level background.

\begin{figure}                      % Fig 4 
\centerline{\psfig{figure=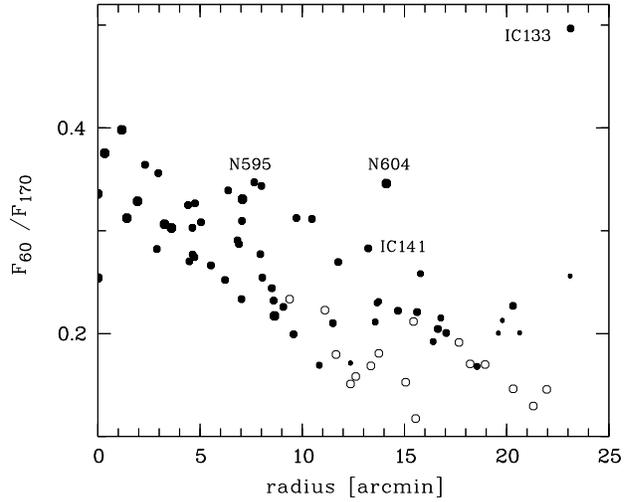,clip=t,width=8.2cm}} 
\caption[]{The $F_{60}/\,F_{170}$ ratio versus deprojected distance from the 
galaxy center. Large dots are for star-forming regions with high 100$\mu$m flux 
density, small dots for faint ones. Some prominent regions are labeled by their names. 
Empty circles stand for inter-arm regions. } 
\end{figure}

\subsection{Correlation of localized 60\,${\mu}$m emission with H$\alpha$ emission} % 4.2

The location of the FIR knots is also traced in the distribution of H\,{\sc ii} 
regions, best seen in the optical, in the light of the H$\alpha$ line. In order 
to correlate the dust emission heated by the UV radiation from star-forming 
regions with the H$\alpha$ emission from the associated H\,{\sc ii} regions, we 
make use of the map in Fig. 6. Before, the H$\alpha$ map needs to be convolved with 
a Gaussian profile of 60\arcsec\ FWHM to achieve a similar spatial resolution. 
Fig. 6 shows the central 33\arcmin$\times$33\arcmin\ area of this map. Its 
morphology is almost indistinguishable from the map of localised warm dust emission 
at 60$\mu$m, $F_{60}^l$ (Fig. 5), except for the fact that the ratio of FIR to 
$H_{\alpha}$ emission appears to be higher in the central area than in the 
outskirts. 

\begin{figure}                        % Fig 5 
\centerline{\psfig{figure=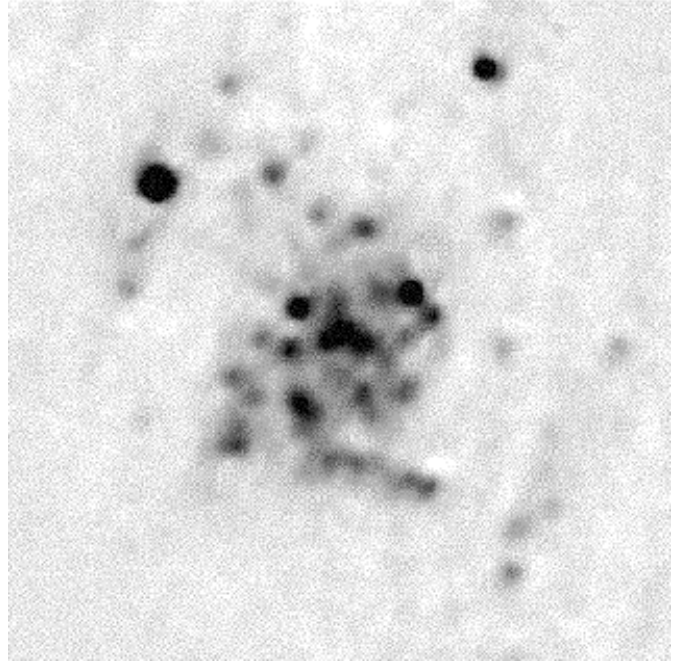,clip=t,width=8.8cm}} 
\caption[]{Distribution of the localised warm dust component at 60\,$\mu$m, 
$F_{60}^l$, visualised by the scaled difference map $2(F_{60}-0.165\times\,F_{170})$, 
with the factor 0.165 given by the average flux density ratio  $F_{60}/F_{170}$ in the 
inter-arm regions. and the factor 2 given by Eq. 1.} 
\end{figure}

\begin{figure}                         % Fig 6 
\centerline{\psfig{figure=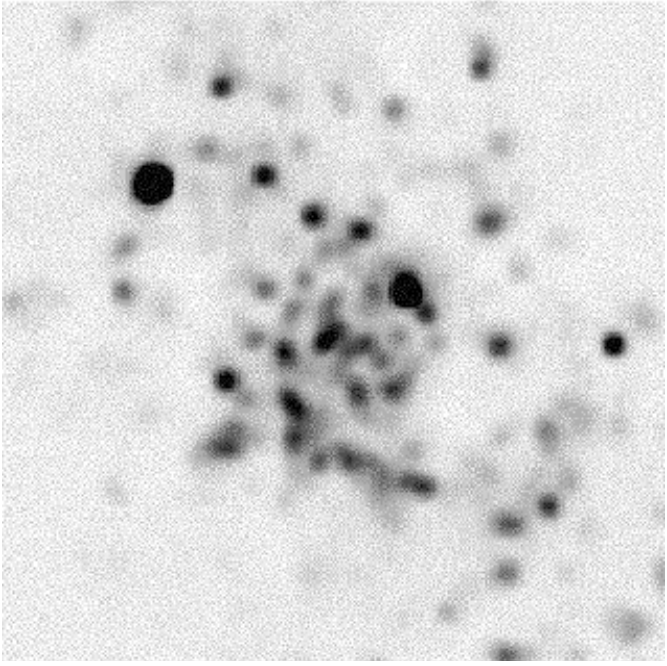,clip=t,width=8.8cm}} 
\caption{H$\alpha$ map of M33 convolved to a resolution of 60\arcsec} 
\end{figure}

The $F_{\rm H\alpha}/F_{60}^{l}$ ratio (Fig. 7, where $F_{\rm H\alpha}$ is 
the flux in the line) for the FIR knots shows a clear systematic increase 
with increasing radial distance from the center (allowing for the [NII] line 
contribution decreasing with distance, the slope would be even steeper). The 
low data point at $r=15$\,arcmin in Fig. 7 originates again from the star-forming 
region IC133 with its unusually high 60$\mu$m flux density value. 
The least squares fit to the data points provides a slope of 
d(log($F_{H\alpha}/F_{60}^{l}$))/d($r$)\,=\,0.038/arcmin.

A significant increase of the $F_{\rm H\alpha}/F_{\rm FIR}$ ratio towards the outer 
parts of the galaxy was also reported by Devereux et al. (1997). Their slope is however 
less steep, since it includes not only the star-forming regions but also locations 
inbetween the spiral arms, where the ratio is systematically lower.

\begin{figure}                            % Fig 7 
\centerline{\psfig{figure=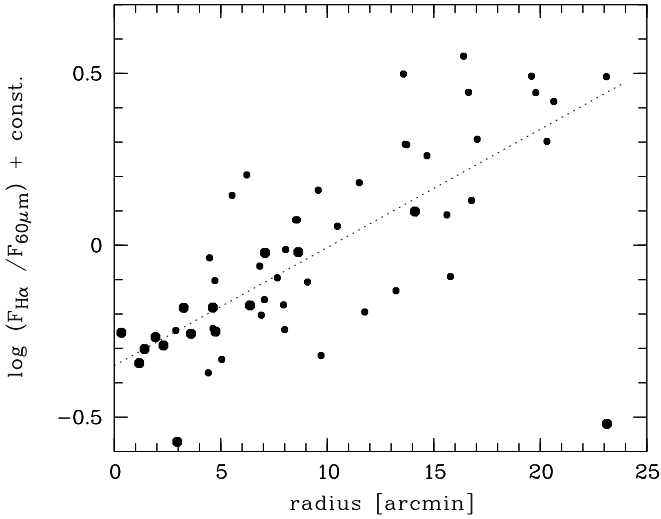,clip=t,width=8.8cm}} 
\caption{$F_{H\alpha}/F_{60}^{l}$ ratio for star-forming regions versus distance from 
the galaxy center $radius$. Symbol sizes indicate the brightness of the sources. 
(The ordinate is scaled in such a way that the mean ratio is 1)} 
\end{figure}

\subsection{Extinction}                  % 4.3

While the foreground extinction towards M\,33 is negligible ($A_V$ = 0.09mag, de 
Vaucouleurs et al. 1991, or $A_V$ = 0.18mag, Burstein \& Heiles 1982 and 
Schlegel et al. 1998), Israel \& Kennicutt (1980) and Berkhuijsen (1983) derived 
from a comparison of H$\alpha$ data with radio flux densities at 21 and 6.2\,cm 
extinction values of up to 2\,mag. They also found a clear decrease of the extinction 
for the H\,{\sc ii} regions of M\,33 with increasing distance from the galaxy center 
of the form of $A_V=2.3-0.07~r$\,mag (with $r$ in arcmin). 

From a pixelwise comparison of the H$\alpha$ emission with a 6\,cm radio map 
Devereux et al. (1997) found no such trend of the internal extinction with 
radius but a scatter around a typical value of the order of $A_V=1$. They therefore 
assigned the increase of the $F_{H\alpha}/F_{FIR}$ ratio to a metallicity gradient 
in the sense, that hotter and more massive stars occur preferentially in the outer, 
metal-poorer regions of the galaxy. 
It seems that the radius dependence holds for the H\,{\sc ii} regions, but not for the 
diffuse emission, since, by means of multiwavelength observations of Balmer and Paschen 
emission lines in several H\,{\sc ii} regions, Petersen \& Gammelgaard (1997) again 
derived extinction values decreasing with distance from the galaxy center in a similar 
way as found by Israel \& Kennicutt (1980) and by Berkhuijsen (1983). 

When correcting the H$\alpha$ fluxes according to the above $A_V$ - $r$ relation, 
the radial increase in $F_{\rm H\alpha}/F_{60}^l$ mostly disappears, suggesting that 
this increase is mainly due to varying extinction. This suggestion is supported 
by the relation between extinction and cold dust radiation discussed in the next 
section. 

There might be other processes contributing to the trend shown in Fig.\,7, such 
as a dust depletion at locations of low gas density, or/and effects due to metallicity 
gradients (Devereux et al. 1997).

\subsection{The cold dust}            % 4.4

Analogous to the isolation of the localized emission component at 60 micron 
(Sect. 4.1), we now construct a map of the diffuse cold dust component by 
suppressing from the 170$\mu$m map the contribution from the star-forming complexes: 
\begin{eqnarray} 
F_{170}^{d} = \frac{F_{170} - F_{60}/f^{l}}{1 -f^{d}/f^{l}} 
\end{eqnarray} 
The scaled difference map is shown in Fig. 8. 

Since the $F_{60}/F_{170}$ ratio is higher than average in the bright knots (see Fig. 4), 
the difference map is overcorrected at these locations, thus yielding holes. 
Apart from these defects one can see that the spiral structure has almost disappeared. 
Instead, a disk like feature with an E-W extension of $\sim25$\arcmin\ becomes apparent, 
revealing the distribution of the diffuse cold dust emission. 
The 170$\mu$m-bright spiral arm segment in the South coincides spatially with a region 
of very strong CO emission (Lequeux, priv.comm.), probably a region of forthcoming 
violent star formation. 

The original 170$\mu$m map shows that large amounts of dust are located in or 
around the 
star-forming regions. This cold dust associated with the massive molecular cloud 
complexes is expected to weaken the optical line emission of the embedded 
H\,{\sc ii} regions. A comparison between the $F_{\rm H\alpha}/F_{60}^l$ and 
the 170$\mu$m flux density shows (with the exception of NGC\,604) that indeed 
the brighter 170$\mu$m sources exhibit lower $F_{\rm H\alpha}/F_{60}$ ratios. 
Between the center of M33 and a deprojected radial distance of 15\arcmin\ the 
$F_{\rm H\alpha}/F_{60}^l$ ratio 
increases by a factor of about 3.3, while the surface brightness S$_{170}$ 
(averaged over annuli) decreases by about 120\,MJy\,sr$^{-1}$ over the same distance. 
Assuming half of the cold dust to be located in front of the H\,{\sc ii} regions, 
and taking this dust to be responsible for the reddening and the variation of 
$F_{\rm H\alpha}/F_{60}^l$, this gradient would imply 
\begin{eqnarray} 
E_{\rm B-V} \sim 0.01 \times S_{170} \hspace{3mm} {\rm [MJy\,sr}^{-1}] 
\end{eqnarray} 

% fact=4 is eq. to 1.5mag 
% AHa = 1.5mag/120/2MJy/sr -> AV=1.8mag/60MJy/sr 
%$A_V$~[mag] $\sim 0.03~F_{170}$~[MJy\,sr$^{-1}$] 
%$E({\rm B-V))$~[mag] $\sim 0.02~F_{100}$~[MJy\,sr$^{-1}] 

Schlegel et al. (1998) derived from galactic cirrus extinction a similar 
correlation between E(B-V) and the cold dust emission, ~ 
$E_{\rm B-V}=0.018\times DT$~, 
where $DT$ represents the IRAS 100$\mu$m flux density corrected to a reference 
temperature of 18.2K using the DIRBE/IRAS maps.

\begin{figure}                        % Fig 8 
\centerline{\psfig{figure=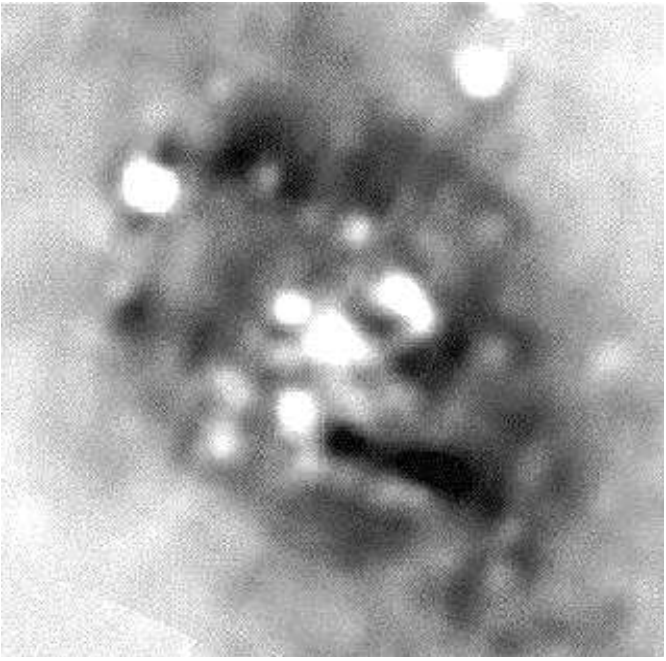,clip=t,width=8.8cm}} 
\caption[]{Distribution of the diffuse cold dust component at 170\,$\mu$m, 
$F_{170}^d$, given by the scaled difference map $2(F_{170} - F_{60}/0.32)$. } 
\end{figure} 

The supposition that the intrinsic extinction is traced by the cold dust would 
give a straightforward explanation for the apparent contradiction between 
Devereux et al. (1997) who find no gradient in their pixel-wise comparison of 
$F_{H\alpha}$ and $F_{6cm}$ (their figure 7) and the gradient for the H\,{\sc ii} 
regions presented by Israel \& Kennicutt (1980), since the cold dust is 
concentrated towards the star-forming regions as well. 

The cold dust constitutes a large amount of mass in galaxies (e.g. Klaas et al. 
2001, Popescu et al. 2002). Assuming an average grain size of 0.1$\mu$m and a 
density of 3\,g\,cm$^{-3}$, the cold dust emission in M33 is equivalent to a mass 
of $6\times10^6$\,M$_{\odot}$. This is about 20\% of what is found in M\,31 
(Haas et al. 1998), which is expected considering the higher mass of the latter. 

% extrapolated S850=56Jy,  BB850(16K)=??   kappa850=0.865cm2/g,  D=830*3.1*10**21cm 
% Mdust = D**2/kappa S850/BB850 =6.2e6 Msun (nach Klaas-Programm) 

\subsection{Star formation}                % 4.5

In the following paragraphs we will discuss how far the FIR and optical star 
formation calibrators lead to consistent results for the whole galaxy as well 
as for the individual star-forming regions. 

An integration of the two blackbody fit for M33 (see Fig. 3) over the wavelength 
range 40 to 250$\mu$m, thus including the strong diffuse cold dust component, 
leads to 7.6~10$^{-11}$\,W\,m$^{-2}$ = 1.5 10$^9$\,L$_{\odot}$ for M33. 
This is almost two times the value that one would derive from our 60 and 
100\,$\mu$m points using Helou et al.'s (1988) relation 
$F_{40-120}=1.26~(2.58 F_{60}~+~F_{100})~10^{-14}$ [W\,m$^{-2}$], 
with $F_{60}$ and $F_{100}$ in Jy, which do not include the 
cold dust component. 

Kennicutt (1998) derived from the star-burst synthesis models of Leitherer 
\& Heckman (1995), for solar abundances, Salpeter IMF, and assuming that 
the dust re-radiates all the bolometric luminosity, a relation between 
star-forming rate and L$_{\rm FIR}$ of 
1.7~$10^{-10}~L_{\rm FIR}$\,M$_{\odot}$\,yr$^{-1}$\,L$_{\odot}^{-1}$. 
Since the percentage of re-radiated light for normal galaxies is only around 
30\% (Popescu \& Tuffs 2002b), the calibration factor used for starburst 
galaxies will probably underestimate the SFR in M33. 
Hughes et al. (1998) derived for high redshifted HDF galaxies a calibration 
factor of 2.3~10$^{-10}~L_{FIR}$ M$_{\odot}$\,yr$^{-1}$\,L$_{\odot}^{-1}$, 
resulting in a star-forming rate of 0.35~M$_{\odot}$\,yr$^{-1}$ for M33, 
while the calibration of Rowan-Robinson et al. (1997) of 
4.5~10$^{-10}~L_{\rm FIR}$ M$_{\odot}$\,yr$^{-1}$\,L$_{\odot}^{-1}$ 
yields a value of 0.69~M$_{\odot}$\,yr$^{-1}$, and a recent empirical relation 
between $L_{FIR}$ and star-forming rate (Bell 2003) gives 0.33~M$_{\odot}$\,yr$^{-1}$. 

Applying the $F_{H\alpha}$-$SFR$ calibration given by Kennicutt (1998) to the total 
H$\alpha$ line luminosity of M\,33, and correcting it for a mean $A_V=1.2$, (Israel 
\& Kennicutt 1980, Devereux et al. 1997) and for a mean [N\,{\sc ii}] contribution 
(averaged over the galaxy) of 20\% (McCall et al. 1985, Vilchez et al. 1988), 
the star-forming rate results 
to $SFR($H$\alpha$)\,=\,0.47~M$_{\odot}$\,yr$^{-1}$, which is in satisfactory 
agreement with the FIR values. 

When we, however, pick a substructure of the galaxy - for instance NGC\,604, 
the brightest H\,{\sc ii} region in M\,33 - no consistency can be achieved between 
the SFRs derived from the optical and from the FIR. 
For this example, the H$\alpha$ luminosity 
corrected for an intrinsic extinction of $A_V=0.8$ (the mean of Israel \& 
Kennicutt 1980, and of Peterson \& Gammelgaard 1997), and for 
17\% [N\,{\sc ii}] 
contribution for NGC\,604 (Vilchez et al. 1988), yields a star-forming rate of 
0.046\,M$_{\odot}$\,yr$^{-1}$. By comparison, if all the UV light were 
locally absorbed, its observed FIR luminosity of 5.8 10$^7$\,L$_{\odot}$ would 
yield SFR values of only between 0.010 (Kennicutt 1998) and 0.026\,M$_{\odot}$\,yr$^{-1}$ 
(Rowan-Robinson et al. 1997), just 40\% 
of the value derived from the optical. Thus, in this case, 
some 60\% of the stellar UV light escapes the dust sphere around the 
H\,{\sc ii} region and contributes to the heating of the diffuse dust 
in the galaxy disk!

% The distance modulus for M33 is 8.3e45 W/(W/m2)  (for D=830kpc) or 2.11e19 Lsun/(Wm2) 
% 
% M33 total: F(Ha)= 3.4e-13W/m2 (Devereux): (L(Ha) = 7.06e+6) = 2.8e33 W 
% = 7.1e6Lsun (wie Devereux, Table 2) 
% for 25% [NII] and mean extinction for H{\sc II} regions AV~1.2mag AHa=1.0mag: 5.6e33 
% equivalent to SFR(Ha)=0.45 Msun/yr. 
% FIR= 2.05a-14*(1.36*525+0.96*1180+0.63*2240) = 2.05e-14*(764+1152+1411)=6.6e-11 W/m2 
% = 1.4e+9Lsun. 
% 
% Line flux of NGC604 is 400e-16 W/m2, extincion A(Ha)~0.62 mag, [NII] contribution 25\% 
% Thus, we have a corrected line flux of 400e-16*2.5**0.65*0.75*8.3e45= 4.4e+32 W (11.3e6Lsun), 
% equivalent to $SFR = 0.035 M_{\odot}$\,yr$^{-1}$. 
% from the FIR we get $F_{FIR} = 2.05e-14*(1.36*30 + 0.96*46 + 0.63*83) = 2.8e-12$ W/m2, or 
% $L_{FIR}= 5.8e7Lsun, 
% Umrechnungsfaktor ist  1.0W/m2 = 2.11e19Lsun 

\subsection{A local radio-FIR correlation} 

\begin{figure} [b]                           % 9 
\centerline{\psfig{figure=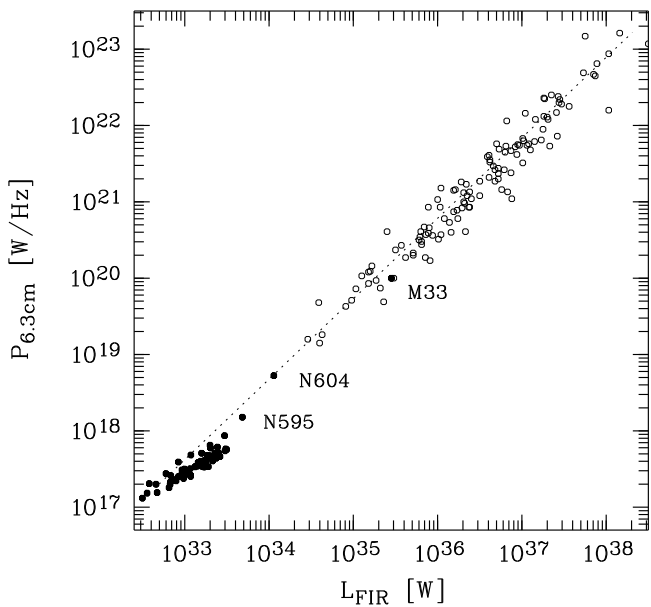,clip=t,width=8.8cm}} 
\caption[]{Plot of the monochromatic radio luminosity $versus$ the FIR luminosities 
for M33 and its star-forming regions (filled circles) together with the data for 
the Effelsberg 100-m galaxy sample (Wunderlich et al. 1987, open circles). The 
dotted line has a slope of 1.10 (Wunderlich \& Klein 1988).} 
\end{figure}

One of the tightest correlations in extragalactic astronomy is that between the 
FIR luminosity, as calculated from the IRAS flux densities at 60 and 100 $\mu$m (for 
example combined in the above-mentioned relation $F_{40-120}$ by Helou et al. 
1988), and the differential radio continuum luminosity in the GHz range for 
late-type galaxies. It was originally found by de Jong et al. (1985), Helou et 
al. (1985) and Wunderlich et al. (1987), and holds for the spatially integrated 
emissions of objects whose luminosity is dominated by star formation and not by 
an active nucleus. 

This radio-FIR correlation can be theoretically explained, since the Interstellar 
Medium in such galaxies acts as a global calorimeter for the relativistic 
electrons and, in large part, for the UV photons, both of which are 
generated by the stars 
(V\"olk 1989). Usually the correlation is found 
to be somewhat steeper than unity, the radio luminosity increasing faster than 
linearly with the FIR luminosity. Restricting the FIR to the IRAS 60 $\mu$m 
luminosity, Yun et al. (2001) have recently obtained a nearly linear 
correlation, based on a very large sample of 1809 galaxies from the local 
Universe, which Bell (2003) considers as a matter of conspiracy of different 
conflicting effects. 

Our present work allows us to investigate whether such a correlation also exists for 
the typically sub-kpc size galactic substructures such as the star-forming 
regions in M33. For this purpose we made use of the 6.3\,cm radio map of 
Buczilowski \& Beck (1987), which has a spatial resolution of 2.4\arcmin. 
Since this resolution is considerably lower than the resolution of our 60 and 
100$\mu$m maps we smoothed the FIR maps to a comparable spatial resolution, 
before rederiving the photometry for the knots in M33. 
For the estimate of the FIR fluxes we used the relation given by Helou et al. 
(1988) to allow a comparison with IRAS data. 
Fig. 9 shows the radio flux plotted versus the FIR flux. 
There appears to be indeed a correlation. Even taking into account the difficulties 
inherent in the data reduction, the slope of this correlation is smaller than unity, 
of the order of 0.9. 
We can ask ourselves why we should expect such a ``local'' correlation. 

A priori this is not obvious. The absorption of the non-ionizing stellar UV 
photons depends strongly on the dust column density through the star-forming regions. And even 
for NGC 604 about 60 percent of the UV light escapes the dust sphere of the region, 
as mentioned above. For smaller regions this fraction is expected to be higher. The 
rest goes into the diffuse dust in the galactic disk. 
This is the dominant FIR part in the standard global radio-FIR correlation, 
but it can be neglected in the ``local'' correlation, 
unless the surface filling factor of the ensemble 
of regions becomes a significant fraction, say more than 10 percent of the disk 
area. Nevertheless, the fact that we did not include the 170\,$\mu$m in the FIR 
luminosity of Fig. 9 emphasizes the shorter wavelengths, and we noted earlier 
that a large fraction (about 80\%) of the emission at 60\,$\mu$m comes from the 
localized component. Thus, some fraction, say 30\%, of the 
spatially integrated IRAS FIR luminosity may come from individual star-forming 
regions and will be proportional to their respective star 
formation rates. In fact, a higher star formation rate will probably imply a 
higher dust column density and thus a stronger than linear dependence of the FIR 
luminosity on the local star formation rate. 

Let us next consider the radio emission from individual regions. If we assume that 
the sources of the radio emitting relativistic electrons are mainly the remnants 
of core collapse supernovae, then their progenitor stars will have remained 
in the star-forming region 
during their evolution. In addition to this non-thermal radiation, 
the progenitor stars will give rise to local thermal (free-free) 
radio emission. Therefore the necessary condition for a ``local'' 
correlation is satisfied, because the sources for both emissions are localized in 
the star-forming region. 

Also, for a face-on galaxy, the star-forming regions are detected along the direction 
perpendicular to the disk. Along that direction we see the integrated emission, 
and the electrons confined to this column will have lost their energy by Inverse 
Compton and synchrotron emission, and thus their luminosity can be proportional 
to the local star formation rate. Much more important is the following effect: the 
supernova remnants which accelerate the electrons, confine them during the 
active remnant life time of $\sim 10^5$~yr, during which time they radiate 
with a very hard energy spectrum. Acceleration automatically implies 
confinement, because {\it in the shock frame} the particles are convected in 
the downstream direction. During their confinement the electrons radiate in the 
interior effective magnetic field which can be expected to be several times, say 
5 times higher than in the average interstellar medium (Bell \& Lucek 2001; 
Berezhko et al. 2002). In agreement with observations (Biermann 1976, Ulvestad 
1982) this implies a ratio of $\sim\,(25\times 10^5$\,yr$)/(3\times 10^7$\,yr$) \sim 10^{-1}$ 
between the synchrotron luminosity from the ensemble of star-forming regions to the diffuse 
luminosity from the overall galaxy where the particles have a life time of $\sim 
3\times 10^7$~yrs. Nevertheless, most of the relativistic electrons will leave 
the region after their release from the sources before having lost their energy 
radiatively. This is easily seen, since their radiative loss length 
$(\kappa(E)\times \tau_{loss}(E))^{1/2}  \geq$ few kpc\footnote{
Here $\kappa(E)$ is the spatial diffusion coefficient of electrons
of kinetic energy $E$, and $\tau_{loss}(E)$ is the electron energy loss time.
The product $\kappa(E)\times \tau_{loss}(E) \propto E^{0.1}$ is almost
energy-independent.} exceeds the typical sizes of star-forming regions, and this 
is even a lower limit (Ptuskin et al. 1997). Therefore, like the cold 
dust emission from Fig. 8, the diffuse synchrotron emission will constitute a 
featureless disk-like structure and can be neglected in the local correlation.  
In fact, the escape from a star-forming region will be the more 
effective the higher the cold gas mass is there, and should therefore decrease 
with increasing FIR luminosity, if anything. 

Therefore, we must expect some local radio-FIR correlation in star-forming regions, and even one 
which has a slope $< 1$. This explains the observational results shown in Fig. 9. 
However, this correlation is primarily due to the confinement of synchrotron 
electrons in their sources (supplemented by local free-free emission), and due 
to local dust absorption. As such it will primarily be seen through the warm 
dust emission observed by IRAS. For the 
total FIR as measured by ISOPHOT we expect a much weaker correlation, if at all. 
Quantitatively, the correlation depends very much on the local transport 
conditions for photons and relativistic electrons and has little connection with 
the global radio-FIR correlation because the ingredients of the latter 
correlation can be neglected in the local correlation. 
In this picture one would predict that the radio emission responsible 
for the local correlation will have a hard spectrum with a spectral 
index flatter than about 0.5.

\section{Conclusions.}           % 5

The FIR radiation from M\,33 is composed of a warm ($\sim 45$K) and a 
cold ($\sim 17$K) dust emission component, 
which were separated by fitting the SEDs of the entire galaxy and of 
individual regions 
with two modified Planck functions of emissivity $\beta \propto \lambda^{-2}$. 
We spatially isolated the two dust emission 
components and proved that they can be identified with separate morphological 
components. 
The warm dust is preferentially associated with the star-forming regions. 
The cold dust arrises in part from a diffuse disk of about 25\arcmin\ 
diameter, but also has a localized component 
associated with the molecular clouds in star-forming regions. 

Estimates of the star formation rate derived from the FIR for the whole galaxy 
are broadly consistent with that derived from the H$\alpha$ emission. In the 
case of individual star-forming regions, a consistency between rates derived 
from the optical and from the FIR allows 
only a fraction of the UV radiation to be absorbed locally. 

The correlation between the $F_{H\alpha}/F_{60}$ ratio and the 170\,$\mu$m 
surface brightness is in agreement with the extinction relation established 
by IRAS/DIRBE observations for the cirrus in the Milky Way. 

The ISOPHOT maps were also used to derive a local radio-FIR correlation for 
star-forming regions in M~33. It is argued that this local correlation is 
due to quite different emission components than to those that lead to 
the well-known global radio-FIR correlation.

\acknowledgements{
It is a pleasure to thank Dr. Hermann-Josef R\"oser for carrying out the optical observations, 
and Drs. Rainer Beck, Eric Bell and Cristina Popescu for stimulating discussions. 
The ISOPHOT Data Center at MPIA are supported by Deutsches Zentrum f\"ur Luft- und Raumfahrt 
with funds of Bundesministerium f\"ur Bildung und Forschung, grant no. 50 QI 0201. }


\begin{thebibliography}{} 
\bibitem[]{} Acosta-Pulido, J.A., Gabriel, C., \& Casteneda, H.O., 2000, Experi\-mental 
             Astronomy, 10, Kluwer Academic Publishers, p333 
\bibitem[]{} Bell, E. 2003, ApJ 586, 794 
\bibitem[]{} Bell, A.R., \& Lucek, S.G. 2001, MNRAS, 321, 433 
\bibitem[]{} Berezhko, E.G., Ksenofontov, L.T., \& V\"olk, H.J. 2002, A\&A, 395, 943 
\bibitem[]{} Berkhuijsen, E.M. 1983, A\&A, 127, 395 
\bibitem[]{} Biermann, P. 1976, A\&A, 53, 295 
\bibitem[]{} Buczilowski, U.R., \& Beck, R. 1987, A\&ASS, 68, 171 
\bibitem[]{} Burstein, D., \& Heiles, C. 1982, AJ, 87, 1165 
\bibitem[]{} Chini, R., Kreysa, E., Kr\"ugel, E., \& Mezger, G. 1986, A\&A, 166, L8 
\bibitem[]{} de Jong, T., Klein., U., Wielebinski, \& R., Wunderlich, E. 1985, A\&A, 147, L6 
\bibitem[]{} de Vaucouleurs, G., de Vaucouleurs, A., Corwin, H. G. Jr., et al. 
             1991, Third Reference Catalogue of Bright Galaxies (Springer). 
\bibitem[]{} Devereux, N.A., Duric, N., \& Scowen, P.A. 1997, AJ, 113, 236 
\bibitem[]{} Gabriel, C., Acosta-Pulido, J., \& Heinrichsen, I. 1998, Astronomical Data 
             Analysis Software and Systems VII, A.S.P. Conference Series, Vol. 145, 165 
\bibitem[]{} Haas, M., Lemke, D., Stickel, M., et al.  1998, A\&A, 338, L33 
\bibitem[]{} Helou, G., Soifer, B.T., \& Rowan-Robinson, M. 1985, ApJ, 298, L7 
\bibitem[]{} Helou, G., Khan, I.R., Mallek, L., \& Boehmer, L. 1988, ApJS, 68, 151 
\bibitem[]{} Hippelein, H.H. 1986, A\&A, 160, 374 
\bibitem[]{} Hughes, D.H., Serjeant, S., Dunlop, J., et al.  1998, Nature, 394, 241 
\bibitem[]{} Israel, F.P., \& Kennicutt, R.C. 1980, ApLett, 21, 1  % Star forming laws 
\bibitem[]{} Kennicutt, R.C. 1998, ApJ, 498, 541  % Star forming laws 
\bibitem[]{} Kessler, M.F., Steinz, J.A., Anderegg, M.E., et al. 1996, A\&A, 315, L27 
\bibitem[]{} Klaas, U., Haas, M., M\"uller, S.A.H., et al. 2001, A\&A, 379, 823 
\bibitem[]{} Laureijs, R.J., Klaas, U., Richards, P.J., et al. 2000, The ISO Handbook, 
             Vol. V 
\bibitem[]{} Leitherer, C., \& Heckman, T.M.  1995, ApJS, 96, 9 
\bibitem[]{} Lemke, D., Klaas, U., Abolins, J., et al. 1996, A\&A, 315, L64 
\bibitem[]{} McCall, M.L., Rybski, P.M., \& Shields, G.A. 1985, ApJS, 57, 1 
\bibitem[]{} Odenwald, S., Newmark, J., \& Smoot, G.  1998, ApJ, 500, 554 
\bibitem[]{} Petersen, L., \& Gammelgaard, P.  1997, A\&A 323, 697
\bibitem[]{} Popescu, C.C., Misiriotis, A., Kylafis, N.D., et al. 2000, A\&A, 362, 138 
\bibitem[]{} Popescu, C.C., Tuffs, R.J., V\"olk, H.J., et al. 2002, ApJ, 567, 221 
\bibitem[]{} Popescu, C.C., \& Tuffs, R.J., 2002a, Reviews in Modern 
             Astronomy 15: Astronomy with Large Telescopes from Ground and Space, 
             ed. R.E. Schielicke, Jena, Astronomische Gesellschaft, p239 
\bibitem[]{} Popescu, C.C., \& Tuffs, R.J. 2002b, MNRAS, 335, L41 
\bibitem[]{} Ptuskin, V.S., V\"olk, H.J., Zirakashvili, V.N., \& Breitschwerdt, D. 
             1997, A\&A, 321, 434 
\bibitem[]{} Rice, W., Boulanger, F., Viallefond, F., Soifer, B.T., \& Freedman, W.L. 
             1990, ApJ, 358, 418
\bibitem[]{} Rowan-Robinson, M., Mann, R.G., Oliver, S.J., et al. 1997, MNRAS, 289, 490 
\bibitem[]{} Schlegel, D.J., Finkbeiner, D.P., \& Davie, M. 1998, ApJ, 500, 525 
\bibitem[]{} Schmidtobreick, L., Haas, M., \& Lemke, D. 2000, A\&A, 363, 917 
\bibitem[]{} Schulman, E. \& Bregman, J.N.  1995, ApJ, 441, 568 
\bibitem[]{} Stickel, M., Lemke, D., Klaas, U., et al. 2000, A\&A, 359, 865 
\bibitem[]{} Tuffs, R.J., \& Popescu, C.C. 2002, in 
             Exploiting the ISO Data Archive. Infrared Astronomy in the Internet Age, 
             ed. C. Gry et al., ESA Publications Series, ESA SP-511, in press 
\bibitem[]{} Tuffs, R. J., Popescu, C. C., Pierini, et al. 2002, ApJS 139, 37 
\bibitem[]{} Tully, R.B. 1988, {\it Nearby Galaxies Catalog}, Cambridge Univ. Press 
\bibitem[]{} Ulvestad, J.S. 1982, ApJ, 259, 96 
\bibitem[]{} Vilchez, J.M., Pagel, B.E.J., Diaz, A.I., et al. 1988, MNRAS, 235, 633
\bibitem[]{} V\"olk, H.J. 1989, A\&A, 218, 67 
\bibitem[]{} Wunderlich, E., Klein, U., \& Wielebinski, R.  1987, A\&AS, 69, 487 
\bibitem[]{} Wunderlich, E. \& Klein, U.  1988, A\&A, 206, 47 
\bibitem[]{} Yun, M.G., Reddy, N.A., \& Condon, J.J.  2001, ApJ, 554, 803 
\end{thebibliography}
\end{document}